\documentstyle[prl,aps,twocolumn]{revtex}
\begin{document}
\draft
\twocolumn[\hsize\textwidth\columnwidth\hsize\csname
@twocolumnfalse\endcsname

to be published {\it Modern Physics Letters B}

\title{Polaron Dissociation at the Insulator-to-Metal Transition}
\author{P. Qu\'emerais and S. Fratini }
\address{ Laboratoire d'Etudes des Propri\'et\'es Electroniques des
Solides,\\ (LEPES/CNRS), BP 166, 38042 Grenoble Cedex 9, France\\ }
\bigskip
\date{\today}
\maketitle
\begin{abstract}

Considering the long range Coulomb interactions between large polarons in
dielectrics, we propose a model for their crystallization when no bipolarons
are formed. As the density increases, the melting is examined at $T=OK$. One
possibility is the delocalization towards a liquid state of polarons. However,
we show that this cannot happen if the electron-phonon coupling is larger
than some critical value. The other competing mechanism is the dissociation
of the polarons themselves, favored owing to their large mass at strong
coupling. Finally, we propose a phase diagram for the insulator-to metal
transition as a function of the density and electron-phonon coupling.

\bigskip

\end{abstract}

]

\bigskip

\narrowtext

Any theory of band structure for interacting fermions is based on the fact
that the long range part of the Coulomb repulsion is screened and contained
in collective modes, called plasmons~\cite{bohm}. However, for an idealized
low density system the long range Coulomb repulsion dominates the kinetic
energy, and Wigner~\cite{wigner} has shown that the carriers crystallize in
that case. This does not occur for electrons in usual metals, owing to the
high density of carriers.

For doped insulators, the carriers' density induced by doping can be extremely
low. These materials can be classified into two types: polar and non-polar.
Non-polar compounds are the usual covalent semiconductors which are easily
doped up to the insulator-to-metal transition. Such materials are
characterized by their dielectric constant $\kappa$, which does not depend on
the frequency (the static dielectric constant $\varepsilon_s$ and the high
frequency dielectric constant $\varepsilon_{\infty}$ are close together
$\kappa \approx \varepsilon_s \approx \varepsilon_{\infty}$). The
insulator-to-metal transition is due to the screening and ionization of the
localized impurity states (the effects of disorder are neglected here).
Mott~\cite{mott} has given a criterium to evaluate the critical density above
which the ground-state is metallic. Given the radius   $R_{imp.}=\hbar^2
\kappa /e^2 m^*$ of the impurity bound state ($m^*$ is the effective mass of
the carrier), the system is  metallic for doping densities $n$ such that $n >
n_c^M$, where $(n_c^M)^{1/3} R_{imp.} \approx 0.25$. Mott assumes that the
long range Coulomb interactions  between liberated carriers are also well
screened as soon as $n>n_c$. This is correct because at these densities the
kinetic energy of the carriers is much higher than the Coulomb energy $
1/{\kappa} R_s$ ($R_s$ is related to the density by $1/n=4 \pi R_s^3/3$),
which is reduced by $\kappa$ (which can be large), so that we are sure that
the Wigner localization would occur only at densities much lower than the
critical Mott density.

For polar insulators, such as some transition metal oxides, the situation is
more complicated owing to the formation of polarons. When the static
($\varepsilon_s$) and high frequency ($\varepsilon_{\infty}$) dielectric
constants have an appreciable difference of magnitude, a free carrier is
bound to the polarization that it creates around itself. The couple made of
the carrier with its surrounding lattice polarization is called a
polaron~\cite{landau,frohlich}. For the single polaron problem, one defines
the dimensionless electron-phonon coupling $\alpha =\left({{{m^*} / {2\hbar
^3\omega _{ir}}}}\right)^{1/2}{{e^2} / {\tilde \varepsilon }}$. The constant
$\tilde \varepsilon$, which is responsible for the polaron formation, is
defined as $1/ \tilde \varepsilon = 1/ \varepsilon_{\infty}-1/\varepsilon_s$,
and $\omega_{ir}$ is the phonon frequency of the optical longitudinal mode.
The polaron problem has been extensively addressed in many theroretical
studies~\cite{kuper}. However, Feynman~\cite{feynman} has proposed a
variational method using path integral calculations giving with a very good
accuracy the  ground state properties for any value of $\alpha$. The idea is
to use a simplified variational Hamiltonian to calculate quantities of
interest:
						\begin{equation}
						H_0={{p^2} \over {2m^*}}+{{P^2} \over {2M}}+ {K \over 2}
						(x-X)^2 \label{H_0} .
						\end{equation} 
Here the electron with coordinate $x$ and momentum $p$, is coupled to a
fictitious particle with coordinate $X$ and momentum $P$ which simulates the
polarization field. $K$ and $M$ are the variational parameters. As pointed
out by Feynman, eq.(\ref{H_0}) well imitates the polaron. In particular, $M$
gives a good estimate to the true effective mass of the polaron : $M_P
\approx m^*+M$, and the polaron radius can also be defined as $R_P \approx
\sqrt{\hbar /2M_P{\left({K/M_P}\right)^{1/2}}}$. Table 1 gives the rest mass
$M_P$ obtained by Feynman and Schultz~\cite{feynman} for different values of
$\alpha$. We notice that for large $\alpha$, the polaron mass is much larger
than the mass of the free carrier $m^*$.

If one considers now two polarons in the system, each electron (or hole) is
attracted by the polarization induced by the other. On the other hand, there
is also a direct repulsion between them. The resulting competition yields two
different situations depending on the ratio $\eta = \varepsilon_{\infty} /
\varepsilon_s$. Many theoretical calculations~\cite{vinetski} have shown that
if $\eta < \eta_c \approx 0.1$, a bound state between two polarons is formed,
which is called a dielectric bipolaron. If this condition is not satisfied,
the Coulomb repulsion dominates and both polarons repel as $1/ \varepsilon_s
d$, where $d$ is their relative distance. We will assume from now on that we
are in this last situation, which is often verified in real materials. All
along this letter, we will consider materials which verify $0.1< \eta <0.5$.

On chemically doping at finite density polar dielectrics with $\eta > 0.1$,
we are faced with three intricated problems: (i) the polaron formation, (ii)
their binding to the impurity atoms (known to be the F-Center
problem~\cite{pekar}), and (iii) the Coulomb repulsion between polarons if
they are mobile. As already pointed out by one of us~\cite{queque}, the last
point (iii) is essential to understand the physics of such systems. Since the
polaron mass can be large (Table 1), the kinetic energy of a set of polarons
which behaves as $ 1/M_P R_s^2$ can remain small compared to the Coulomb
energy $1/ \varepsilon_s R_s$ even at high density and high dielectric
constant. To simplify the problem and focus mainly on the effects of the
long range Coulomb interactions, we shall concentrate in this letter on an
idealized system of polarons at finite density and interacting in a jellium of
equal density (and opposite charge). At low density, they
crystallize~\cite{queque} as the electrons in the Wigner crystal. We will now
address the problem of the melting of such a crystal as the density
increases.

The strong electron-phonon coupling limit $\alpha \rightarrow \infty$, or
equivalently the low phonon-frequency limit $\omega_{ir} \rightarrow 0$ can
be first examined. It is well-known~\cite{landau,pekar} that in this limit,
the polaron is formed by an electron localized in a quasi-static polarization
potential-well, which behaves as $-e^2/ {\tilde \varepsilon}r$ at large
distance, and is almost constant at short distance. The mass $M_P$ of the
polaron becomes infinite as $\omega_{ir}$ vanishes, while its energy $E_P$
and radius $R_P$, which are finite, only depend on the dielectric constants and
effective mass $m^*$ of the electron. When the density increases, the
polarons are unable to delocalize owing to their infinite mass. Thus, at some
critical density, the electronic states localized within the
polarization-wells must disappear for the same reason as the localized
impurity states at the usual Mott transition: they are screened by liberated
electrons. Making the same assumptions as Mott in the case of impurity
states, we propose the following qualitative criterium for the dissociation
of the polarons,  valid for  $\omega_{ir} \approx 0$, i.e $\alpha
\rightarrow \infty$:
						\begin{equation} \label{mod_Mott}
						{\left({n_c^P}\right)}^{1/3} \left ( {\tilde \varepsilon \over
						\varepsilon}_{\infty} \right ) R_P \approx Const.
						\end{equation}
The ratio $\tilde \varepsilon / \varepsilon_{\infty}$ takes into account the
different dielectric constants appearing respectively in the calculation of
the Thomas-Fermi length ($\varepsilon_{\infty}$) and in the polaron radius
($\tilde \varepsilon$). The resulting state occuring for $n>n_c^P$ will be
discussed later on. Numerically, on taking the values $\varepsilon_s = 30$,
$\varepsilon_{\infty} =5$, $m^*=m_e$ and $Const =0.25$ \cite{nota}, one obtains $n_c
\approx 10^{19} {cm}^{-3}$.

The above consideration of the insulator-to-metal transition based on the
polaron dissociation is {\it a priori} no more valid at finite electron-phonon
coupling, since in that case the polarons could delocalize to form a liquid
state instead of being dissociated. These two competing mechanisms for the
insulator-to-metal transition are now more precisely examined at finite
electron-phonon coupling and zero temperature. To modelize the
situation at finite polaron density, we add to the host material
a density $n$ of electrons plus a {\it rigid} compensating jellium 
of same density.
To account for the strong Coulomb interactions in a kind of mean-field
approximation, 
starting from the low density limit, we take as a simplified model,
a single electron localized within a uniformly positively charged and {\it
polarizable} sphere of radius $R_s$, which is nothing but
the Wigner crystal of polarons (see ref.~\cite{queque} for details). As done by Wigner for the
electron crystal, we neglect the dipole-dipole interactions between adjacent spheres. The
hamiltonian describing such a system is then given by:
				\begin{eqnarray}
					H &=&{{p_e^2} \over {2m^*}}-\int\limits_{space} {{\bf P}_{ir}\cdot \left({\bf
					D}^- + {\bf D}^+\right)}
				+{1 \over {4 \pi \varepsilon_{\infty}}}\int\limits_{space}{{\bf D}^- \cdot
				{\bf D}^+} \nonumber\\
				&+& {1 \over 8 \pi \varepsilon_{\infty}}\int\limits_{space}{{\bf D}^+\cdot
				{\bf D}^+} + {{2\pi \tilde \varepsilon } \over{\omega
				_{ir}^2}}\int\limits_{space} {\left( {\dot P_{ir}^2+\omega
				_{ir}^2P_{ir}^2} \right)} ,
				\end{eqnarray} 
where ${\bf D}^+$ and ${\bf D}^-$ are the electric displacement due to the
jellium and to electron respectively. ${\bf{P}}_{ir}$ is the polarization. We
now introduce the polarization part which only responds to the electron
motion : ${\bf{\tilde P}}_{ir} = {\bf P}_{ir}-(1/4 \pi \tilde \varepsilon)
{\bf D}^+$ and $\dot {\bf {\tilde P}}_{ir}= \dot {\bf P}_{ir}$, and easily
obtain:
					\begin{eqnarray}
					H &=& C_0 + {{p^2} \over {2m^*}}+{1 \over 2} m^* \omega_{W}^2
           x^2 \nonumber\\
					&-& \int\limits_{space} {\bf {\tilde P}}_{ir} \cdot { {\bf D}^-}
					+ {{2\pi \tilde \varepsilon } \over{\omega
				 _{ir}^2}}\int\limits_{space} {\left( {\dot {\tilde P}_{ir}^2+\omega
				 _{ir}^2 {\tilde P}_{ir}^2} \right)} \label{H_pol_wig} .
				 \end{eqnarray}   
$C_0= - 9e^2/10 \varepsilon_s R_s$ is a constant part of the energy at fixed
density, and $\omega_W^2 = e^2 / m^* \varepsilon_s R_s^3$
is related to the usual plasmon frequency $\omega_p^2 = 4 \pi n e^2/m^*$
by the relation $\omega_W^2 = \omega_p^2 / 3 \varepsilon_s$.
Equation (\ref{H_pol_wig}) is nothing but the hamiltonian of a polaron, the
electron of which is bound in a harmonic potential. Taking the limit $R_s \rightarrow \infty$, one
recovers the usual single polaron hamiltonian, with $C_0=\omega_W =0$, ${\bf D^+}=0$, and
${\bf{\tilde P}}_{ir} = {\bf P}_{ir}$.

As previously proposed by Nozi\`eres {\it et al.}~\cite{nozieres,maradudin},
the Lindemann criterium allows a good qualitative discussion for the melting
of the electron Wigner crystal. The essential idea is that the crystallized
state is unstable when the quantum fluctuations (zero point motion) of the
electron positions exceed some critical value. More precisely, the crystal
melts when $\left < {\delta x^2} \right >^{1/2} / R_s > \delta$, where
$\delta$ is a phenomenological constant usually taken as $\delta \approx
1/4$. For the Wigner crystal,  $\left < {\delta x^2}\right > = \hbar/2m^*
\left({e^2/m^* R_s^3}\right)^{1/2}$, and the critical value of $R_s$ below
which the crystal melts~\cite{comment} is\cite{nozieres}  : $R_c^W=64 a_0$ 
($a_0$ is the Bohr radius).

The application of this criterium to the polaron crystal is more complicated
owing to the composite nature of the polaron. To be able to delocalize the
polarons towards a liquid state, the quantum fluctuations for each single
polaron  moving as a whole, i.e. the motion of the electron together with the
surrounding polarization,  must become large in comparison with $R_s$.	On the
contrary, if the relative fluctuations of the electron with respect to the
polarization become large, the polaron dissociation is favored. The Feynman
model yields a natural way to evaluate the behavior of such quantities: since
the polarization field is replaced by a particle with coordinate $X$, one
easily calculates both the quantum fluctuations of the center of mass
$R=(m^*x+MX)/(m^*+M)$ and the relative coordinate $r=x-X$. Thus, we have now
two different Lindemann criteria:

(i) ${\left < \delta R^2 \right >}^{1/2}/R_s>1/4$ which allows the melting of
the crystal towards a liquid state of polarons ;

(ii) ${\left < \delta r^2 \right>}^{1/2}/R_s>1/4$ for the polaron
dissociation.

Both quantities are calculated by the same variational treatment, just taking
instead of equation (2), the new Hamiltonian: $H_0^W=H_0 + \gamma x^2 /2$,
where $\gamma$, $K$ and $M$ are the new adjustable parameters. Once the
lattice degrees of freedom are integrated out~\cite{feynman}, the action for
the model (\ref{H_pol_wig}) is given by:
	   \begin{eqnarray}
    S &=& \frac{m^*}{2}\int dt \, \dot{x}^2(t) +
          \frac{m^*\omega_W^2}{2}\int  dt \,  x^2(t) \nonumber \\
      &-& \frac{\alpha}{\sqrt{8}}\int\int dt \, ds \, 
          \frac{e^{-|t-s|}}{ |x(t)-x(s)| } ,
    \end{eqnarray} 
where the integrals are calculated over a time interval $T$, $\hbar=1$ and
the frequency unit is $\omega_{ir}$. The trial (quadratic) action
corresponding to $H_0^W$ is now  
	   \begin{eqnarray}
    S_0^W &=& \frac{m^*}{2}\int dt \,  \dot{x}^2(t) +
              \frac{\gamma}{2}\int dt \,  x^2(t) \nonumber \\
          &+& \frac{C}{2} \int\int dt \, ds \,  [x(t)-x(s)]^2
              e^{-w|t-s|}  ,
    \end{eqnarray}
where $w^2=K/M$ and $C=Kw/4$.

At low density, the two degrees of freedom $r$ and $R$ are uncoupled,
and their frequencies are respectively
$\omega_{int}$ and $\omega_{ext}$.
These are sketched in Fig. \ref{frequencies} as functions of $n$.
 The two corresponding
quantum fluctuations are then given by $\delta r \approx \left( \hbar /2m^* \omega_{int}
\right)^{1/2}$ and $\delta R \approx \left( \hbar /2 M_P \omega_{ext} \right))^{1/2}$. 
Fig.\ref{raggi_massa}.(a) shows the numerical values of the ratios (i) and (ii). One observes that
for fixed $\alpha$, while $\langle \delta r^2 \rangle^{1/2}/R_s$ increases with increasing
density, $\langle \delta R^2 \rangle^{1/2}/R_s$ is bounded from above. In addition, its maximum
value decreases with increasing $\alpha$, so that above a certain critical value $\alpha^*$, the
value $1/4$ is never reached and the crystal melting is prevented.

The situation encountered at infinite electron-phonon
coupling (eq.(2)) is thus continued down to finite values of $\alpha$. For
$\alpha> \alpha^*$, {\it the liquid state of polarons cannot exist}. Physically, $\omega_{ext.}$
being limited by the frequency of the phonons themselves ($\omega_{ext.}\le \omega_{ir}$, see Fig.
\ref{frequencies}), our result means that, when the electron-phonon coupling  is too large, the
polarization {\it cannot follow} the increase in kinetic energy induced by the doping through
$\omega_{W}$. The quantum fluctuations are then transferred to the {\it internal} degree of
freedom. For that reason, the polarons dissociate at the transition.

Let us focus more precisely on the shape of $\langle \delta R^2 \rangle^{1/2}/R_s$ in Fig.
\ref{raggi_massa}(a).  Starting at low density ($R_s \rightarrow \infty$,
$\omega_W\approx 0$), the  extra term $m^*\omega_W^2 x_e^2 /2$ is a
perturbation with respect to the single-polaron energy, and the polaron
properties are almost unchanged as long as its radius is much smaller than the
average distance between carriers.  To a first approximation, we can then
neglect the  composite nature of the  polaron and treat it as a rigid particle
of mass $M_P$ (the single polaron mass from Table 1) moving in a weak 
harmonic potential with a frequency \mbox{$\omega_{ext}=\omega_W
(m^*/M_P)^{1/2} $}.  The average quantum fluctuations in the jellium potential
are given by the following expression
	   \begin{equation} \label{R_low}
    \langle \delta R^2 \rangle^{1/2} = \sqrt{\frac{\hbar}{2M_P (\omega_W
    \sqrt{m^*/M_P})}}  ,
    \end{equation}
which fits very well the low density part of $\langle \delta R^2
\rangle^{1/2}/R_s$ (right side of Fig.  \ref{raggi_massa}.(a)). This gives
the critical $R_c^{\left( i \right)}$ corresponding to the first Lindemann
criterium (i) in terms of \mbox{$R_c^W=64 a_0$} 
    \begin{equation} \label{Rcrit}
    R_c^{\left( i \right)}=R_c^W \frac{\epsilon_s}{\left( M_P/m^* \right)} .
    \end{equation}
If the mass of the carriers is strong enough, despite the static screening of
the charges, the crystallisation is favored in  comparison with the ordinary
Wigner lattice. On the other hand, $\langle \delta r^2 \rangle^{1/2}$ in (ii)
is nothing but the polaron radius, so that (ii) gives : $R_c^{\left( ii
\right)}=4 R_P$.

At higher densities, the frequency of  vibration of the polaron
$\omega_{ext}=\omega_W (m^*/M_P)^{1/2}$ increases,  due to the tightening of
the jellium potential. Thus, if the density is such that
$\omega_{ext} \approx \omega_{ir}$, i.e.  $\omega_W \approx \omega_{ir} (M_P/m^*)^{1/2}$, the
quenching of the polaron frequency is translated into an enhancement of the polaron mass, which
can be described through a function $M_P(R_S)$. This is observed numerically, as shown in Fig.
\ref{raggi_massa}.(b) (the rest mass is recovered for $R_s \rightarrow \infty$). As a consequence,
at high density, the quantum fluctuations of the polaron will be $(\hbar/2M_P(R_s)
\omega_{ir})^{1/2}$ for the left part in Fig. \ref{raggi_massa}.(a), while eq.(\ref{R_low}) is
valid for the right part. This gives rise to the  bell shape of $\langle \delta R^2
\rangle^{1/2}/R_s$. This allows a simple evaluation of $\alpha^*$: the maximum
value  of $\langle \delta R^2 \rangle^{1/2}/R_s$ will be obtained
approximately by evaluating the low density formula eq.(\ref{H_pol_wig})  at 
$\omega_W \approx \omega_{ir} (M_P/m^*)^{1/2}$. If we equate the result to
the melting value $1/4$ we get $(M_P/m^*)^{1/2} \alpha \approx 16
(\epsilon_s/\tilde\epsilon)$. As an example, taking $\epsilon_s=30$ and
$\epsilon_\infty=5$, and using the values for $M_P$ of Table 1, one obtains
$\alpha^*\approx 9$. The variational result, without such approximations,
gives $\alpha^* \simeq 7.5$.

The phase diagram Fig. \ref{diag_fase} gives the calculated critical radii for
crystal melting (i) and polaron dissociation (ii) as  functions of the
electron-phonon coupling $\alpha$.  Though the transition radii decrease
linearly with $\varepsilon_s$ and $\tilde\varepsilon$, the general shape of
the diagram does not change in the studied range $0.1 < \varepsilon_{\infty}
/ \varepsilon_s <0.5$, and so does the critical value $\alpha^*$  ($\alpha^*
\approx 6-9$).

For small $\alpha$ the transition corresponds to the melting of the usual
Wigner crystal of electrons weakly renormalized by the electron-phonon
coupling. The corresponding metallic region ((I) in Fig.2) has been previously
examined by Mahan~\cite{mahan}. Extension of his calculations in the metallic state have
been performed by Devreese {\it et al.} at intermediate coupling, and more recently
by Iadonisi {\it et al.} at strong coupling\cite{mahan}.

For intermediate coupling (say $5< \alpha < \alpha^*$), it can be expected
that the polarons remain well defined, yielding a liquid state of polarons
below the transition (region (II) in Fig.2). The physics in that case may be
well described by the Holstein model for non-interacting
polarons~\cite{holstein,ciuchi}.

For strong coupling ($\alpha > \alpha^*$), the crystal cannot melt towards a
polaron liquid. At the transition, the polarons dissociate and the resulting
metallic state (region (III) in Fig.2)) will present both strong
electron-phonon and electron-electron interactions. A theory for such a system
has been recently proposed by Castellani {\it et al.}\cite{castellani}: it presents phase
separation due to the electron-phonon coupling, which is prevented by the Coulomb repulsion.
The resulting effects are strong charge density wave fluctuations, and a
superconducting instability. Another possibility, already proposed by one of us \cite{queque} is
that only part of the polarons dissociate to quasi-free electrons at the critical density.
In that case, just above the insulator-to-metal transition the system would present  a kind of
electronic pseudo-separation between localized polarons and quasi-free electrons. However, at
higher density, this pseudo-separation would disappear owing to screening effects, and in that case
the system would be well represented by the theory of Castellani{\it et al.}\cite{castellani}.
This possibility will be examined in a further publication.

According to our theory,  the possibility of polaron dissociation  at the
insulator-to-metal transition in real materials depends on the value of 
$\alpha$, which is generally lower than $\alpha^* \approx 6-9$. However, any
mechanism which increases the polaron mass $M_P$ compared to the present 3D
case, will shift $\alpha^*$ to lower values and make the dissociation of
polarons physically reasonable. Moreover, the critical density for this dissociation will be
increased compared to the present 3D case ($n_c^{3D}
\approx 10^{19} {cm}^{-3}$, see eq. (2)). Two examples of such mechanisms can be
mentioned: (a) a strong anisotropy of the effective mass $m^*$, which corresponds to considering
two dimensional polarons; and (b) the formation of a polaron in a Mott-Hubbard insulator, which is
associated to  a spin-polaron\cite{alexandrov}, which mass increases exponentially with its
size.

In conclusion, we have proposed a model which takes into account in a
self-consistent way the effects of the Coulomb interaction for a system of
polarons at finite densities. A new mechanism for the insulator-to-metal
transition has been put in light, which corresponds to the polaron
dissociation at strong electron-phonon coupling.

This work received financial support from the European Commission (contract
no. ERBFMBICT 961230).

\begin{table}[h]
\begin{tabular}{|c|ccccc|} 
\hline
$\alpha$   &   3    &   5  &   7   &   9   &   11  \\ \hline
$M_P/m^*$  &   1.8  &  3.6 &  13.2 &  59.2 &  181  \\ 
\hline
\end{tabular}
\caption{Polaron effective mass as given by Feynman.}
\end{table}

\begin{figure}
\caption{The two frequencies $\omega_{int.}$ and $\omega_{ext.}$ as functions of
the density are represented in arbitrary units for $\alpha=10$. The external frequency
corresponding to the quantum fluctuations of the polaron as a whole saturates and is necessarily
lower than $\omega_{ir}$. The critical density $n_c$ for the polaron dissociation (see text) is
also quoted.} \label{frequencies}
\end{figure}
\begin{figure} 
\caption{(a)  The ratios  $\langle \delta R^2 \rangle^{1/2}/R_s$ (bold line)
and  $\langle \delta r^2 \rangle^{1/2}/R_s$ (dotted line) versus $R_s/a_o$ at
$\alpha=\alpha^*$  (see text), for $\epsilon_\infty=5$, $\epsilon_s=30$ and $m^*=m_e$. 
(b) The corresponding effective polaron mass $M_P$ in units of $m^*$.}
\label{raggi_massa}
\end{figure}
\begin{figure}
\caption{Phase diagram for $\epsilon_\infty=5$, $\epsilon_s=30$ and $m^*=m_e$.
Crystal melting (bold line) and polaron dissociation (broken line). The dotted
line is the critical radius given by eq. (8).  The arrow marks the
result for the modified Mott criterium eq. (2). 
For regions I,II and III: see text.}  
\label{diag_fase}
\end{figure}


\begin{references}
%
\bibitem{bohm} D. Bohm and D. Pines, Phys. Rev., 85, 2, p.338 (1951)
%
\bibitem{wigner} P. W. Wigner, Phys. Rev., 46, p.1002 (1934)
%
\bibitem{mott} N. F. Mott, Phil. Mag. 6, p. 2897 (1961)
%
\bibitem{landau} L. D. Landau, Phys. Z. Sowjetunion, 3, p.664 (1933)
%
\bibitem{frohlich} H. Frohlich, Adv. in Phys., 3, p.325 (1954)
%
\bibitem{kuper} C. G. Kuper and G. D. Whitfield, {\it Polarons and excitons},
(Oliver and Boyd, Edinburgh and London, 1962)
%
\bibitem{feynman} R. P. Feynman, Phys. Rev., 97, 3, p. 660, (1955) ;
 T. D. Schultz, Phys. Rev., 116, 3, p. 526, (1959)
%
\bibitem{vinetski} V. L. Vinetskii and M. S. Giterman, Soviet. Phys. JETP, 6,
p.1796 (1958) ; G. Verbist, M.A. Smondyrev, F. M. Peeters, and J. T. Devreese, Phys. Rev. B,
45, 5262 (1992) ; F. Bassani, M. Geddo, G. Iadonisi and D. Ninno, Phys. Rev. B, 43,  5296
(1991)
%
\bibitem{pekar} S. I. Pekar, AEC-tr-5575 Phys., State Publishing House of
Technical and Theoritical Literature, Moscow (1961)
%
\bibitem{queque} P. Qu\'emerais, Mod. Phys. Lett. B, 9, 25, p.1665 (1995)
%
\bibitem{nota} The constant $Const.$ should be lower than the constant for the
usual Mott criterium ($\approx 0.25$) owing to the difference of shape
between a polarization potential well and an impurity potential well.
%
\bibitem{nozieres} P. Nozi\'eres and D. Pines, Phys. Rev., 111, 2, P. 442
(1958)
%
\bibitem{maradudin} R.A. Coldwell-Horsfall, A.A. Maradudin,  
J.Math.Phys.1, 395 (1960)
%
\bibitem{comment} the value $R_c^W=64$ overestimates the quantum fluctuations.
See discussion in Ref.~\cite{nozieres}
%
\bibitem{mahan} G. Mahan, in {\it Polarons in Ionic Crystals and Polar
semiconductors}, ed. J. Devreese (1972) ; L.F. Lemmens, J.T. Devreese and F. Brosens, Phys. Stat.
Sol. 82, 439 (1977) ; G. Iadonisi, G. Capone, V. Cataudella and G. De Fillipis, Phys. Rev.
B, 53, 13497 (1996)
%
\bibitem{holstein} T. Holstein, Ann. of Phys., 8, 325 (1959)
%
\bibitem{ciuchi} S. Ciuchi, F. De Pasquale, S. Fratini, D. Feinberg, Phys. Rev. B 56, 4494, (1997)
%
\bibitem{castellani} C. Castellani, C. Di Castro and M. Grilli, Phys. Rev.
Lett., 75, p.4650 (1995) 
%
\bibitem{alexandrov} A. S. Alexandrov and N. F. Mott, Rep. Prog. Phys. 57,
1197 (1994)
%

\end{references}
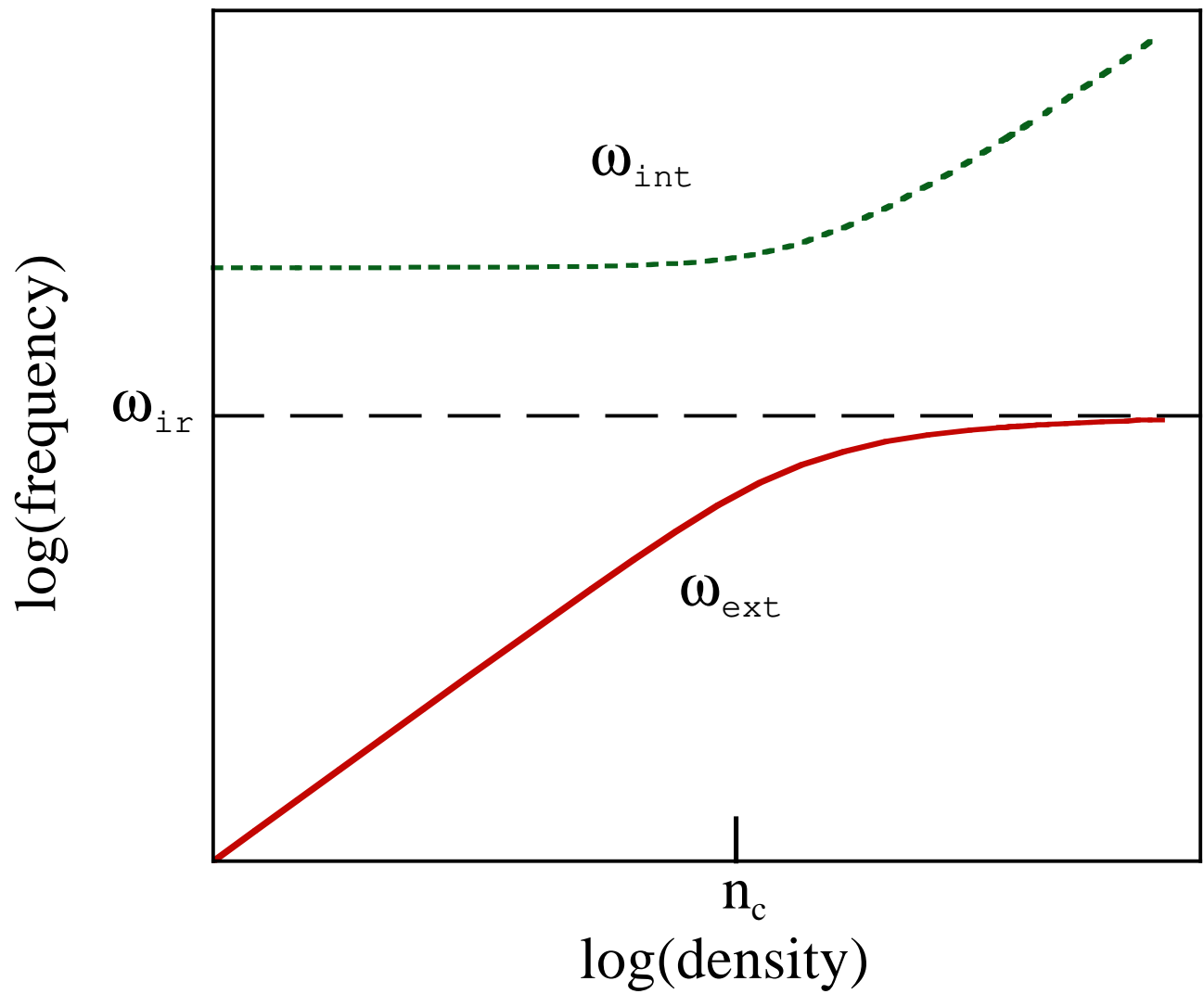
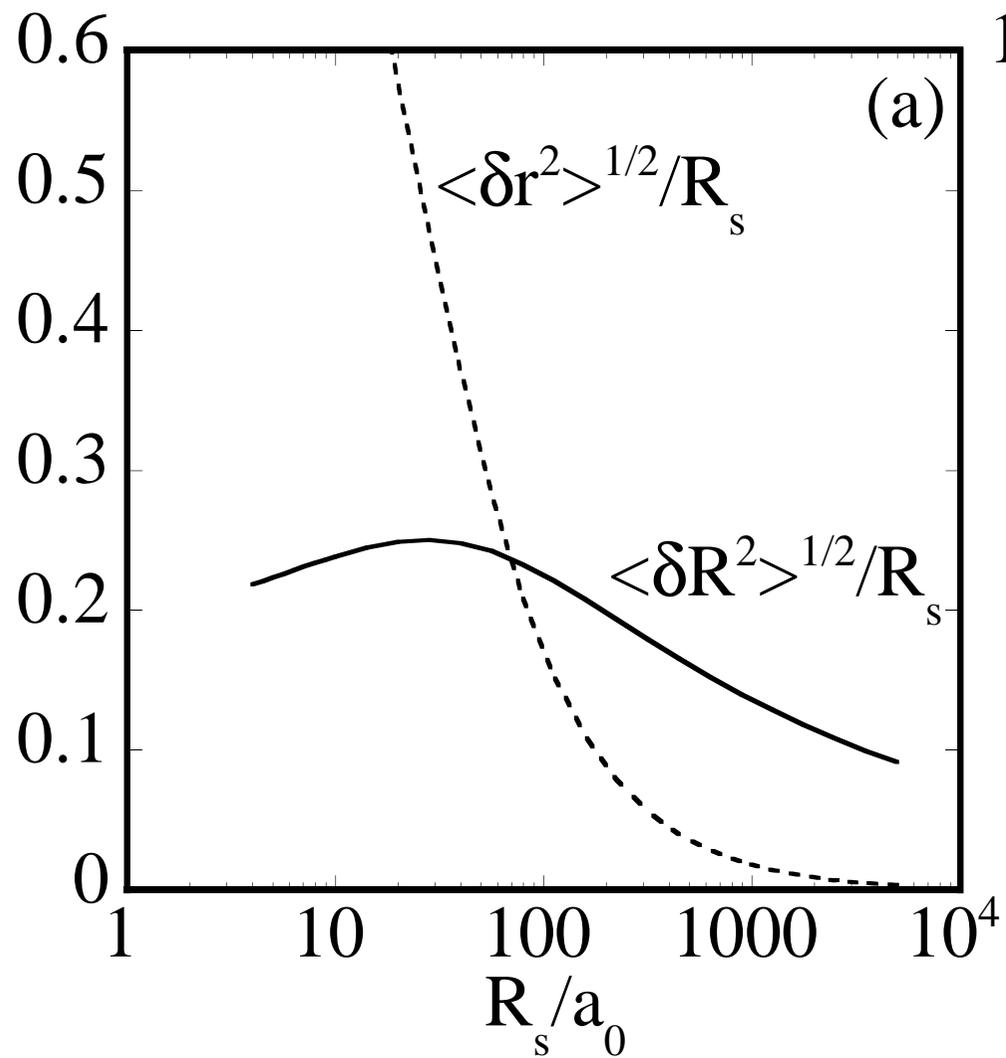
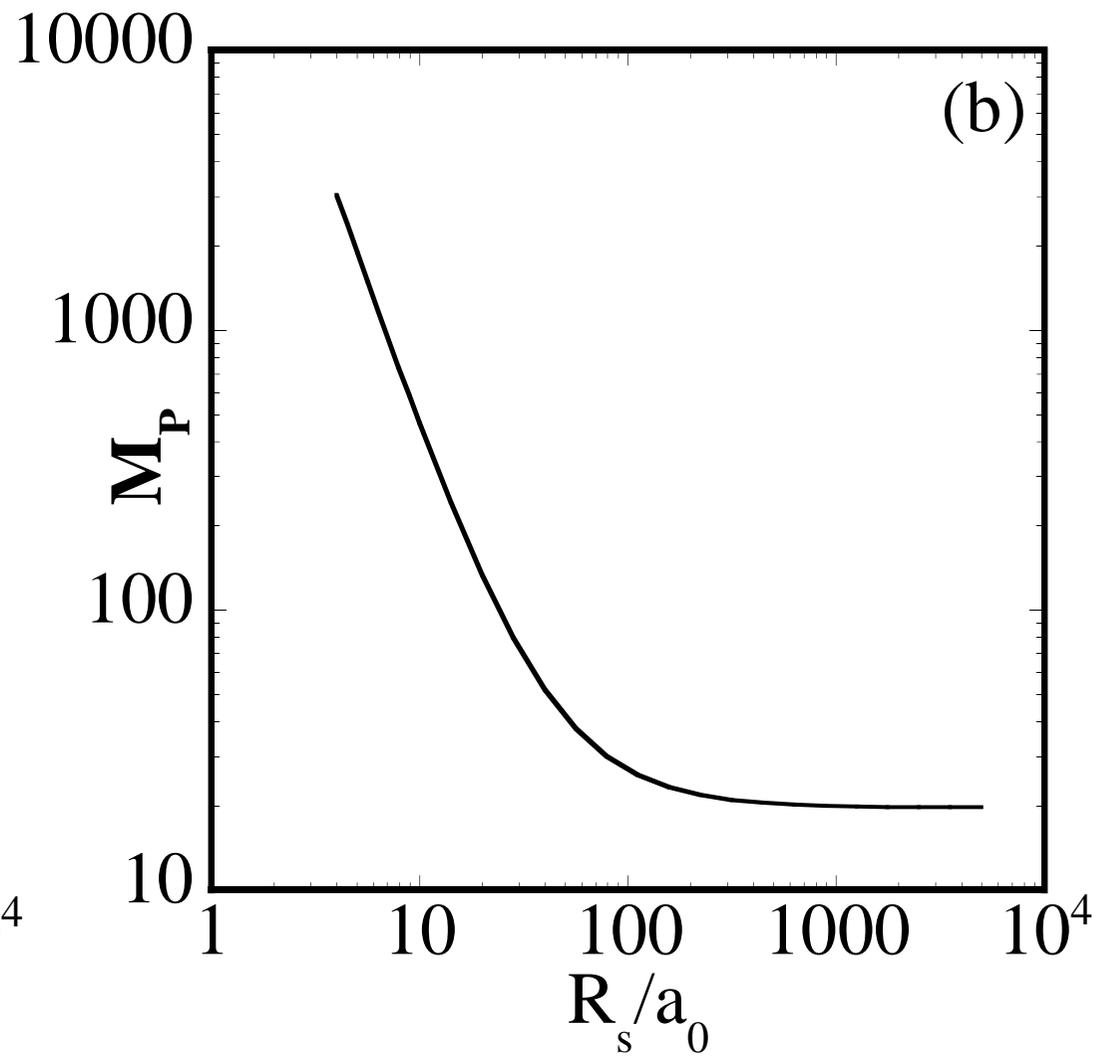
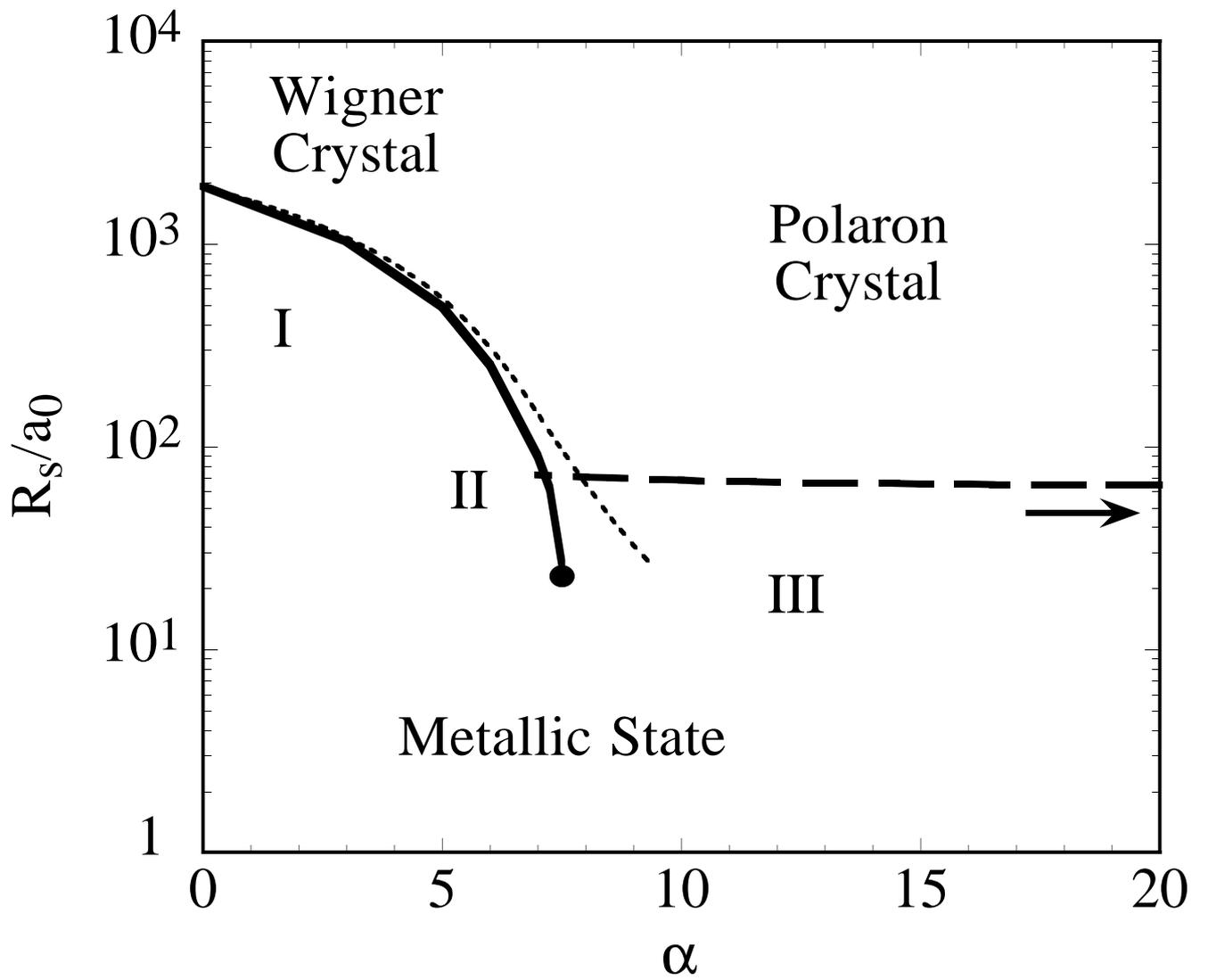
\end{document}